\documentclass[fleqn,usenatbib]{mnras}

\usepackage{newtxtext,newtxmath}
 \usepackage{ulem} 

\usepackage[T1]{fontenc}

\DeclareRobustCommand{\VAN}[3]{#2}
\let\VANthebibliography\thebibliography
\def\thebibliography{\DeclareRobustCommand{\VAN}[3]{##3}\VANthebibliography}

\usepackage{graphicx}	
\usepackage{amsmath}	

\title[Double-Peak Optical Afterglow in GRB 110213A]{Double-Peaked Optical Afterglow in GRB 110213A Inferring a Magnetized Thick Shell Ejecta}

\author[Y. Kusafuka et al.]{
Yo Kusafuka,$^{1}$\thanks{E-mail: kusafuka@icrr.u-tokyo.ac.jp}
Kaori Obayashi,$^{2}$
Katsuaki Asano$^{1}$ and
Ryo Yamazaki$^{2,3}$
\\
$^{1}$Institute for Cosmic Ray Research, The University of Tokyo, 5-1-5 Kashiwanoha, Kashiwa, Chiba 277-8582, Japan\\
$^{2}$Department of Physical Sciences, Aoyama Gakuin University, 5-10-1 Fuchinobe, Sagamihara, Kanagawa 252-5258, Japan\\
$^{3}$Institute of Laser Engineering, Osaka University, 2-6, Yamadaoka, Suita, Osaka 565-0871, Japan
}

\date{Accepted XXX. Received YYY; in original form ZZZ}

\pubyear{\the\year{}}

\begin{document}
\label{firstpage}
\pagerange{\pageref{firstpage}--\pageref{lastpage}}
\maketitle

\begin{abstract}
Gamma-ray bursts early afterglows are important tracers for determining the radial structure and magnetization of the ejecta.
In this paper, we focus on GRB 110213A that shows double-peaked optical afterglow lightcurves and the shallow decay feature of the X-ray afterglow.
We adopt a semi-analytic model for the dynamics of forward and reverse shocks generated through an interaction between an arbitrary magnetized ejecta with a finite thickness and a stratified circumstellar medium. 
Multiwavelength radiation from forward and reverse shocks seen from an arbitrary viewing angle is calculated under a thin-shell approximation. 
Our analysis with multimodal nested sampling methods for GRB 110213A suggests that the thick shell ejecta naturally explains the shallow decay feature of the X-ray afterglow. The combination of the reverse shock emission in the strongly magnetized jet and forward shock emission in the weakly magnetized circumstellar medium makes the double peak feature of the optical afterglows.
The estimated low radiative efficiency in the prompt phase may be a consequence of the high magnetization of the jet in this case.
A multi-messenger emission simulator based on the magnetic bullet afterglow model is publicly available as the open source Julia package "\texttt{Magglow}".
\end{abstract}




\section{Introduction}

Gamma-ray bursts (GRBs) are the most luminous explosions in the Universe. 
Their enormous {isotropic-equivalent} energy $10^{51}\sim10^{55}$ erg is released during a few seconds to minutes, 
followed by a long lasting gradually fading radiation called afterglow. 
Afterglow radiation is generated through the collision of a jet with its surrounding circumstellar medium (CSM). Several theoretical and numerical studies have been widely explored in this external shock scenario and have been applied to constrain the parameters of GRB jets and CSM structures \citep{1992MNRAS.258P..41R,1995ApJ...455L.143S,1997ApJ...476..232M,2005ApJ...628..315Z,2009A&A...494..879M,2020ApJ...896..166R,2025MNRAS.536.1822K,2025MNRAS.540.2098K}. 
Although a flat CSM structure is favoured in most GRBs \citep{2011A&A...526A..23S,2022MNRAS.511.2848A}, a wind-like CSM environment is also proposed \citep{2013NewAR..57..141G,2018ApJ...865...94D,2025arXiv250313291K}. In general, a GRB environment might be more complicated \citep{2024ApJ...962..115R,2025JHEAp..45..392Z,2025MNRAS.540.2098K}, which reflects the uncertain mass loss history of the progenitor star before its explosion. 

The analytical formulae by \citet{1998ApJ...497L..17S} based on the self-similar evolution of the external shock have been frequently used for afterglow lightcurves. In addition, several numerical treatments taking into account the jet geometry \citep[e.g.][]{2012ApJ...749...44V,2020ApJ...896..166R,2022MNRAS.510.1315A} or the evolution of the electron energy distribution \citep[e.g.][]{2017ApJ...844...92F,2020ApJ...905..105A} have been used to constrain the macroscopic and microscopic parameters. 
While a fraction of afterglows can be explained with those monotonically decelerating shocks, $\sim$40\% \citep{2009MNRAS.397.1177E,2018ApJS..236...26L} of early X-ray afterglow lightcurves show shallower decay \citep{2006ApJ...642..389N,2007ApJ...670..565L,2019ApJS..245....1T,2019ApJ...883...97Z} than what the standard model predicts.
Although \citet{2020MNRAS.494.5259Y} point out that GeV/TeV-detected GRBs display no or less prominent shallow decay phase in their X-ray light curves, recently the GeV afterglow has been detected during the X-ray shallow decay phase in GRB 240529A \citep{2025ApJ...986..189T}. 
The energy injection model \citep{2006ApJ...642..354Z,2024ApJ...970..141A} has been considered for the shallow decay phase observed in the early afterglows, 
but the contribution of reverse shock emission may also be necessary to explain the early optical lightcurves \citep[e.g.][]{2006MNRAS.369..197F,2006MNRAS.369.2059P,2009ApJ...702..489R,2014ApJ...781...37P,2014ApJ...785...84J,2019ApJ...879L..26F,2021MNRAS.505.4086G}.

A contribution of a reverse shock emission makes diversity in optical afterglows, characterized by a double-peaked and flattening of the light curves \citep{2003ApJ...595..950Z}. 
Although double-peaked light curves are theoretically expected for most GRBs, there are only a few afterglows detected so far \citep{2014ApJ...785...84J}. 
GRB 110213A is one of the afterglows with double-peaked optical light curves, whose peaks are at $\sim 300$ s and $\sim5000$ s \citep{2011ApJ...736....7C}. This afterglow also shows the shallow decay in their X-ray light curve lasting $\sim 5000$ s, making it difficult to understand within the previous theoretical frameworks. Although \citet{2022ApJ...939...39W} propose the cascade emission origin, both the double-peaked structure and the X-ray shallow decay are not well reproduced. 

Some afterglows are reported to be hard to explain with the conventional models \citep[e.g.][]{2003Natur.424..749G,2006ApJ...638L...5R,2016MNRAS.462.1111D}.
\citet{2025MNRAS.536.1822K} have provided a detailed numerical study of the effects of magnetization and thickness of the ejecta on afterglow lightcurves. 
Higher magnetization makes the reverse shock crossing time shorter, which leads to double-peaked light curves in optical afterglows. The energy injection into the forward shock from the thick shell ejecta weakens the deceleration making shallow decay light curves in X-ray afterglows. 
They also provided a semi-analytic formulae for the dynamics of forward and reverse shocks produced by interactions between a magnetized thick shell ejecta (magnetic bullet) and a homogeneous medium based on their numerical simulation results.  
Assuming the typical particle spectral index of $p=2.2$, the model can produce a gradual flux evolution $F_\nu\propto t_{\rm obs}^{4/3}$ for low frequency and a plateau lightcurve for high frequency $F_\nu\propto t_{\rm obs}^{-0.1}$. 
The contribution of both forward and reverse shock components may produce double-peaked optical light curves with X-ray shallow decay.
In addition, \citet{2025MNRAS.540.2098K} extend the model for the case of a wind-like medium to explain GRB 221009A. 
The methods established in \citet{2025MNRAS.536.1822K,2025MNRAS.540.2098K} 
can be applicable to the complicated afterglows.

In this paper, we apply this method to the double-peaked afterglow in GRB 110213A, especially focusing on the behaviour of the early afterglow emission. 
We develop the semi-analytic treatment for the dynamics of the forward and reverse shock in \citet{2025MNRAS.536.1822K, 2025MNRAS.540.2098K} 
for applying to any stratified medium.
To obtain lightcurves with this methods,
we provide a computational tool "\texttt{Magglow}"\footnote{named after loving sushi: "tuna fish" pronounced "Magglow" in Japanese. The instructions are found in \url{https://github.com/yo3-sun/Magglow}.}: a public, open source Julia package, multi-messenger emission simulator. 
There are several publicly available codes in this field such as \texttt{afterglowpy} \citep{2020ApJ...896..166R}, \texttt{jetsimpy} \citep{2024ApJS..273...17W}, \texttt{PyBlastAfterglow} \citep{2025MNRAS.538.2089N}, \texttt{VegasAfterglow} \citep{2025arXiv250710829W}, etc. However, the propagation effects of the rarefaction wave after the reverse shock crossing have not been included in those codes. \citep[but recently][released an upgrade version of \texttt{jetsimpy} in which includes effects of ejecta thickness]{2025arXiv250715311W}. 
\texttt{Magglow} can calculate the forward and reverse shock radiation from the interaction between arbitrary magnetized ejecta with finite thickness and a stratified circumstellar medium. Although the jet geometry is considered, the current version is limited to only top-hat jets with no lateral spreading. The spreading effects might be important in the late afterglow.

The paper is organised as follows. In Section \ref{sec:model}, we provide semi-analytic formulae implemented in \texttt{Magglow}. In Section \ref{sec:application}, we explain the data analysis about GRB 110213A. In Section \ref{sec:discuss}, we discuss the observational implications of our results for GRB 110213A. The conclusion is summarised in Section \ref{sec:summary}. 

\section{Magnetic Bullet Afterglow} \label{sec:model} 

We describe the semi-analytical model implemented in \texttt{Magglow}. 
This model describes the dynamics of forward and reverse shocks driven by the interaction between a CSM and a magnetized ejecta with isotropic equivalent energy $E_0$, initial Lorentz factor $\Gamma_0$, initial magnetization $\sigma_0$, and initial radial width $\Delta_0$. 
Then, the initial number density of the ejecta $n_{0,\rm ej}$ at the initial radius $R_0$ can be roughly estimated as 
\begin{equation}
    n_{0,\rm ej}=\frac{E_0}{4\pi\Delta_0(1+\sigma_0)\Gamma_0^2m_{\rm p}c^2}R_0^{-2}.
\end{equation}
Since our interest is in the early afterglows, the effects of lateral expansion are neglected \citep[but it is important for the late afterglows, see][for example]{2020ApJ...896..166R}.



\citet{2025MNRAS.536.1822K} describe the dynamics of the forward and reverse shocks of an arbitrary magnetized ejecta with finite thickness propagating in a homogeneous external medium \citep[for the case of wind-like medium, see also][]{2025MNRAS.540.2098K}.
There are three stages of the shock propagation: acceleration, transition, and the Blandford-McKee self-similar stages. The acceleration phase exists only for a high value of $\sigma_0$. The onset of the reverse shock defines the start of the transition phase. In the early phase of the transition phase, the bulk Lorentz factor is almost constant, i.e., the coasting phase. As the reverse shock propagates, the forward shock starts to gradually decelerate, but the deceleration is slower than the Blandford-McKee self-similar solution. This is because the shocked ejecta impedes the deceleration. The crossing of the reverse shock does not directly affect the forward shock. When the influence of the initial shell width disappears, the system turns into the Blandford-McKee self-similar stage. See \citet{2025MNRAS.536.1822K} for the details of the semi-analytical formulae for the shock evolutions.

We extend the model for the case of any stratified medium. The density structure of the CSM as a function of the radius $R$ is given as 
\begin{equation}
    n_{\rm CSM}=n_0(3\times 10^{35})^{k/2} R^{-k},
\end{equation}
where the slope $k$ ranges from $0$ to $3$ \citep{1976PhFl...19.1130B}. 
The dimension of the parameter $n_0$ is $[\mbox{cm}^{k-3}]$.
For the case of $k=0$, $n_0$ simply corresponds to the number density of the interstellar medium. 
For the case of $k=2$, $n_0$ is the same as the conventional wind parameter $A_*=(\dot{M}/10^{-5}\ \rm M_\odot\rm yr^{-1})/(v_{\rm wind}/10^3\ \rm km\ s^{-1})^{-1}~\mbox{cm}^{-1}$, where $\dot{M}$ is the mass loss rate  of the progenitor star with the stellar wind velocity $v_{\rm wind}$ \citep{2012ApJ...758...81M,2018MNRAS.478..110S}.

Initially, the bulk Lorentz factor evolves as $\Gamma \propto t^{1/3}$ for $\sigma_0 \gg 1$. 
Then, the reverse shock is ignited when the magnetic pressure becomes comparable to the pressure in the forward shocked region. The ejecta gradually decelerates after the ignition time, which is obtained with the method in \citet{2025MNRAS.536.1822K}.
The reverse shock crossing time $t_\Delta$ \citep[][]{1995ApJ...455L.143S,2005ApJ...628..315Z} is obtained from
\begin{equation}
    \Delta_0=\int_0^{t_\Delta} c(1-\beta_{\rm RS})\left(1-\frac{\Gamma_{\rm max}n_{\rm ej}}{\Gamma_{\rm RS}n_{\rm RS}}\right)^{-1}dt,
    \label{eq:RS}
\end{equation}
where $\Gamma_{\rm max}$ is the maximum Lorentz factor just after the acceleration phase \citep[see][]{2025MNRAS.536.1822K}, and $n_{\rm ej}$ and $n_{\rm RS}$ are the number density of the unshocked ejecta and that just behind the reverse shock, respectively. The density $n_{\rm RS}$ is calculated with the Rankine-Hugoniot relation for an arbitrary magnetized case \citep{2005ApJ...628..315Z}.  Even after $t_\Delta$, the energy deposited in the shocked ejecta affects the evolution of the shock Lorentz factors. This transition phase lasts until the rarefaction wave catches up with the forward shock front.
The bulk Lorentz factor of the forward and reverse shocked regions in the transition phase evolves as 
\begin{equation}
    \Gamma_{\rm FS}=\Gamma_{\rm RS}=\frac{\Gamma_{\rm max}}{\left[ 1+2\Gamma_{\rm max}\sqrt{\frac{n_{\rm CSM}}{n_{\rm ej}(1+\sigma_{\rm RS})}} \right]^{1/2}},
    \label{eq:Gamma_trans}
\end{equation}
where $\sigma_{\rm RS}$ is the magnetization of the shocked ejecta.  
Since the factor $n_{\rm ej}(1+\sigma_{\rm RS})$ decreases as $\propto t^{-2}$, the asymptotic behaviour of Eq. (\ref{eq:Gamma_trans}) in the transition phase becomes
\begin{equation}
    \Gamma_{\rm FS}\propto t^{-\frac{2-k}{4}}.
\end{equation}

Given the width $\Delta_0$, we can determine the time when the rarefaction wave catches up to the forward shock front as explained in \citet{2025MNRAS.536.1822K}. After that, the Blandford-McKee self-similar solution  $\Gamma_{\rm FS}\beta_{\rm FS}\propto t^{-(3-k)/2}$ is adopted \citep{1976PhFl...19.1130B}.


We consider forward and reverse shock radiation from a top-hat ejecta with finite opening angle $\theta_{\rm j}$ viewed from $\theta_{\rm obs}$ measured from the jet axis. 
Given the Lorentz factor evolutions of forward and reverse shocks, we calculate the evolution of the emission taking into account synchrotron, synchrotron self-Compton (SSC), synchrotron self-absorption (SSA), and $\gamma \gamma$ absorption with the same calculation method described in \citet{2025MNRAS.536.1822K}. 
The cooling processes of the accelerated particle distributions are adiabatic, synchrotron, and SSC cooling, including the Klein-Nishina effects \citep{2009ApJ...703..675N,2010ApJ...712.1232W}. 
The hadronic radiation and cooling processes of pp-collision \citep{2006PhRvD..74c4018K} and photo-meson interaction \citep{2008PhRvD..78c4013K} can be calculated in \texttt{Magglow}, but we neglect any hadronic contributions throughout in this study.
We calculate the radiation flux along the Equal Arrival Time Surface \citep[EATS;][]{1997ApJ...491L..19W,2005ApJ...631.1022G}, adopting the thin-shell approximation. Defining the polar angle $\tilde{\theta}$ from the line of sight (not the jet axis), the observed time $t_{\rm obs}$ at which photons emitted from a position at $(R,\tilde{\theta})$ arrive at the observer is written as
\begin{equation}
    t_{\rm obs}\equiv(1+z)\left[t-\frac{R(t)-R_0}{c}\cos\tilde{\theta}\right],
    \label{eq:t_obs}
\end{equation}
where $z$ is a cosmological redshift and $t$ is time measured in the engine rest frame. 
As the emissivity is uniform on the jet front surface, given $\tilde{\theta}$, the effective factor coming from the integration of the azimuthal angle $\tilde{\phi}$ around the line of sight is calculated analytically \citep[e.g.,][] {2024MNRAS.528.2066P} as
\begin{flalign}
    & \int d\tilde{\phi} _{\rm on-jet}= \nonumber \\
    &\ \ \left\{
    \begin{array}{lll}
        2\pi \ \ \ \ \ \ \ \ \ \ \ \ \ \ \ \ \ \ \ \ \ \ \ \ \ \ \ \ \ \ \ \ \ \ \ \ \ \ \ \ \ \ \ \ \ \ \ \theta_{\rm obs}\leqq\theta_{\rm j}-\tilde{\theta}\\
    2 \arccos\left(\frac{\cos\theta_{\rm j}-\cos\tilde{\theta}\cos\theta_{\rm obs}}{\sin\tilde{\theta}\sin\theta_{\rm obs}}\right) \ \ \ \ \theta_{\rm j}-\tilde{\theta} \leqq \theta_{\rm obs} \leqq \theta_{\rm j}+\tilde{\theta} \\
    0\ \ \ \ \ \ \ \ \ \ \ \ \ \ \ \ \ \ \ \ \ \ \ \ \ \ \ \ \ \ \ \ \ \ \ \ \ \ \ \ \ \ \ \ \ \ \ \ \ \rm otherwise
    \end{array}
    \right..
    \end{flalign}


The dynamics of the forward and reverse shocks are regulated by 7 macroscopic parameters: $E_0$,  $\Gamma_0$, $\sigma_0$, $\Delta_0$, $n_0$, $k$, and $\theta_{\rm j}$. The radiation from both of the shocks are determined by additional 8 microscopic parameters: 
the fractions of shock internal energy transferred to electrons ($\epsilon_{e, {\rm FS}}$, $\epsilon_{e, {\rm RS}}$), 
the fractions transferred to magnetic fields ($\epsilon_{B, {\rm FS}}$, $\epsilon_{B, {\rm RS}}$), 
the power-law injection indexes of the electron energy distribution ($p_{\rm FS}$, $p_{\rm RS}$), and the number fractions of electrons accelerated into non-thermal distributions ($f_{e, \rm FS}$, $f_{e, \rm RS}$).
Here, the subscripts 'FS' and 'RS' are used to distinguish between the forward and reverse shock parameters, respectively.

\section{Application to GRB 110213A} \label{sec:application}

GRB~110213A was detected by \textit{Swift} Burst Alert Telescope (BAT), with a duration of $T_{90} = 48 \pm 16~{\rm s}$ in the $15$--$350~{\rm keV}$ energy range, classifying it as a long GRB.
The isotropic equivalent gamma-ray energy and spectral peak energy with the source redshift $z = 1.46$ \citep{2011ApJ...743..154C} are estimated as $E_{\gamma, {\rm iso}}^{\rm rest}= 7.2^{+0.1}_{-0.08}\times 10^{52}~{\rm erg}$ and $E_{\rm peak}= 98.4^{+8.6}_{-6.9} ~{\rm keV}$, respectively. 

\subsection{X-ray data}

The X-ray data are obtained from the Swift team's website \citep{2007A&A...469..379E,2009MNRAS.397.1177E}, which provides the temporal evolution of the integrated energy flux in the 0.3-10keV band as well as the spectral hardness. The spectral hardness shows five distinct phases over time; we use the data from the second phase onward. The photon index in each phase is as follows:
$1.735^{+0.120}_{-0.088}$ from 147~s to 1432~s,
$1.81^{+0.09}_{-0.05}$ from 1432~s to 8549~s,
$1.95^{+0.12}_{-0.12}$ from 11640~s to 14238~s, and
$1.98^{+0.12}_{-0.11}$ from 17510~s to 580102~s, respectively.
Adopting these values, we calculate the energy flux density at $10~{\rm keV}$ (corresponding to $\sim2.42\times10^{18}~{\rm Hz}$) in each time interval.

\subsection{Infrared, optical, UV data}

The infrared, optical, and ultraviolet datasets are obtained from \citet{2011ApJ...743..154C}.
The extinction values in the $z$-, $r$-, $u$-, $V$-, and $I$-bands are calculated as follows.
The Galactic extinction toward the direction of GRB~110213A is estimated to be  
$A_z^{\rm MW} = 0.411\pm0.041~{\rm mag}$,
$A_r^{\rm MW} = 0.744\pm0.074~{\rm mag}$,  
$A_V^{\rm MW}=0.892\pm0.089~{\rm mag}$, 
$A_I^{\rm MW} =0.490\pm0.049~{\rm mag}$, and
$A_u^{\rm MW} =1.411\pm0.141~{\rm mag}$,
respectively \citep{2011ApJ...737..103S}. 
The rest-frame wavelengths corresponding to these bands are 
$\lambda_z = 0.367~{\rm \mu m}$,
$\lambda_r = 0.254~{\rm \mu m}$, 
$\lambda_V = 0.211~{\rm \mu m}$, 
$\lambda_I = 0.320~{\rm \mu m}$, and
$\lambda_u = 0.141~{\rm \mu m}$, respectively.
The extinction law of the host galaxy is assumed to follow that of the Small Magellanic Cloud ($R_V = 2.93$).
The host galaxy V-band extinction is assumed to be $A_{V({\rm HG})}=0.22\pm0.05~{\rm mag} $\citep{2024ApJ...970...51G}, and the corresponding extinctions in each band are calculated as
$A_z^{\rm HG} = 0.331\pm0.075~{\rm mag}$,
$A_r^{\rm HG} = 0.513\pm0.117~{\rm mag}$,  
$A_V^{\rm HG}=0.659\pm0.075~{\rm mag}$, 
$A_I^{\rm HG} =0.389\pm0.0089~{\rm mag}$, and
$A_u^{\rm HG} =1.05\pm0.243~{\rm mag}$.
Hence, the total extinction in each band is obtained by summing the Galactic and host-galaxy components: 
$A_z = 0.742\pm 0.085~{\rm mag}$,
$A_r = 1.26\pm 0.14~{\rm mag}$,  
$A_V = 1.55 \pm 0.12~{\rm mag}$, 
$A_I = 0.879\pm 0.10~{\rm mag}$, and
$A_u = 2.46\pm 0.28~{\rm mag}$.

\subsection{Spectral and temporal behaviours} 

{As \citet{2025MNRAS.536.1822K} demonstrated, the dual peak structure of optical light curves can be obtained in the case of $\sigma_0\gg1$. 
In this interpretation for GRB 110213A, the first peak $\sim 3\times 10^2$ s corresponds to the reverse shock crossing time, meanwhile the achromatic second peak $\sim 5\times 10^3$ s occurs at the deceleration time of the forward shock (the onset time of the Blandford-McKee self-similar stage).
The exceptionally long deceleration time $\sim 5\times 10^3$ s requires an exceptionally large $E_0\gtrsim10^{55}$ erg, low ISM density $n_0<1$ $\rm cm^{-3}$, or small initial Lorentz factor $\Gamma_0<10^2$ for the thin shell case.
However, the deceleration time can be elongated by an order of magnitude longer for the thick shell case \citep{2025MNRAS.536.1822K}.}

{
For thick shell cases, the dynamics is in the transition phase until $5\times 10^3$ s. 
The rising index is estimated as $(3-p_{\rm FS})/2$ in $\nu_{\rm m}<\nu_{\rm X}<\nu_{\rm c}$ for slow cooling case. For fast cooling case, the index is 0.5 at $\nu_{\rm c}<\nu_{\rm X}<\nu_{\rm m}$ \citep{2025MNRAS.536.1822K}. 
According to the observed X-ray rising index $0.44\pm0.10$ until $3\times 10^2$ s \citep{2011ApJ...736....7C}, the spectral index is roughly estimated as $p_{\rm FS}\sim2.2$ for the slow cooling case. 
After $4\times 10^2$ s, the X-ray lightcurve is almost flat until $5\times10^3$ s, suggesting that the X-ray is in the range of $\nu_{\rm X}>\max(\nu_{\rm m},\nu_{\rm c})$.
In this spectral regime, the decaying index becomes $-(p_{\rm FS}-2)/2$ for both slow and fast cooling cases, indicating $p_{\rm FS}\sim2$.
}

{After the first peak, the decaying index of the optical bands until $2\times10^3$ s are $-1.10\pm0.24$ \citep{2011ApJ...736....7C}, which is shallower than typical decaying index of high latitude emission from reverse shock \citep{2005ApJ...628..315Z}. 
However, a contribution of the gradually rising component of the forward shock emission until $\sim10^3$ s may reproduce the shallow decaying lightcurve. 
After that, the optical fluxes increase gradually until $5\times10^3$ s.
The rising index is expected to be $\sim 0.5$, if the dynamics is in the transition phase.
}

{
The observed decaying index of the optical $(-1.80\pm0.15)$ and X-ray $(-1.98\pm0.07)$ for $10^4\ \rm{s}<t_{\rm obs}<10^6\ \rm{s}$ \citep{2011ApJ...736....7C} are steeper than the value without jet break $(2-3p_{\rm FS})/4=-1.15$ for $p_{\rm FS}=2.2$ \citep{1998ApJ...497L..17S}. 
Taking into account the jet-edge effects after the time when $\theta_{\rm j}<1/\Gamma$ that causes a steepening in the lightcurve by $t_{\rm obs}^{-3/4}$ \citep{1999MNRAS.306L..39M}, the reduced temporal index becomes $-1.15-0.75=-1.9$ close to the observed values. 
}

\subsection{Parameter estimate}

In order to obtain posterior distributions for the model parameters, we employ multimodal nested sampling \citep{2008MNRAS.384..449F}.
Bayesian inference is instrumental in this study for two reasons. First, by deriving the posterior distributions of the parameters, we can directly quantify the degeneracies and uncertainties inherent in the GRB afterglow model, which involves many physical parameters. Second, by computing the Bayesian evidence, we can perform a principled comparison between competing models, such as those assuming an interstellar medium or a stellar-wind environment.
We utilize \texttt{PyMultiNest} \citep{2016ascl.soft06005B}, a Python interface to the MultiNest nested sampling algorithm \citep{2009MNRAS.398.1601F}.
In our analysis, we use 450 initial live points and adopt an evidence tolerance of 0.6 as the convergence criterion.
The likelihood function was assumed to be Gaussian, which corresponds to the assumption that the observational uncertainties are normally distributed with known standard deviations:
\begin{align}
    {\cal L}(\vec{\theta}) 
  = \prod_{k=1}^N \frac{1}{\sqrt{2\pi \sigma_k^2}} \exp \left({-\frac{(Y_k-F_k(\vec{\theta}))^2}{2\sigma_k^2}}\right),
\end{align}
where 
$N$ is the number of observation data points,
$Y_k$ and $\sigma_k$ are the observed flux and standard deviation (1$\sigma$ error) of the $k$-th observed flux ($k=1,2....,N$), respectively, and
$\vec{\theta}$ denotes the model parameters used to compute the theoretical flux $F_k$.

\subsection{Results}

For the GRB 110213A afterglow, the median values of the one-dimensional marginalized posterior distributions, along with the prior ranges for each model parameter, are summarized in Table~\ref{table:parameter_estimation}. 
We assume an on-axis observer ($\theta_{\rm obs} = 0~{\rm rad}$) throughout the analysis. As the ejecta is assumed to be strongly magnetized ($\sigma \gtrsim1$), the magnetic field in the reverse shock region estimated with $\epsilon_{B, {\rm RS}}<1$ is always lower than the value obtained from the ideal MHD condition.
Therefore, we do not need $\epsilon_{B, {\rm RS}}$ as a free parameter.
Log-uniform priors are adopted for $E_0$, $\Gamma_0$, $\sigma_0$, $\Delta_0$, $n_0$, $\epsilon_{e, {\rm FS}}$, $\epsilon_{B, {\rm FS}}$, $f_{e, {\rm FS}}$, $\epsilon_{e, {\rm RS}}$, and $f_{e, {\rm RS}}$, whereas uniform priors are used for $k$, $p_{\rm FS}$, $p_{\rm RS}$, and $\theta_{\rm j}$
(see also Figure~\ref{fig:cornor_plot}). 

Although several parameters reached the edge of their prior ranges and wider priors were also tested, we present the result with the highest Bayesian evidence.
With the best-fit parameter set, we plot the light curves in Figure~\ref{fig:test_LC}. 
The early bump observed at approximately $5 \times 10^2$~s is dominated by the reverse shock emission, while the second bump around $5 \times 10^3$~s is mainly due to forward shock emission.
{
The overall agreement between the model and the data is moderate, with a reduced $\chi^2$ value of $\chi^2/{\rm d.o.f} \sim 1.9$ for the best-fit model (see also Table~\ref{table:parameter_estimation}).
The corresponding $p$-value is close to zero, indicating that the model does not provide a statistically perfect fit.
The main contribution to the reduced goodness of fit comes from the early X-ray excess (see Figure~\ref{fig:test_LC}), where the theoretical light curve remains almost flat, whereas the observed data show a rising trend. This discrepancy is discussed in detail in Section~\ref{subsec:early_phase}.
In addition, the discrepancy in the $z$-band, where the theoretical flux slightly underestimates the observed data, also contributes to the reduced $\chi^2$ value.
However, this deviation is within a factor of $\sim 1.5$, and the model successfully reproduces the overall trend of the light curve, including the observed double-peaked structure.
}

The model demands a large radial width ($\Delta_0/c \sim 10^3$ s) and magnetization $\sigma_0 \sim 10$. To reproduce the unusually long onset time of forward shock emission $\sim 10^3$ s, a large $\Delta_0$ compared to $cT_{90}$ is required to maintain the transition phase longer. Using the best-fit parameters, the flux evolution is $\propto t_{\rm obs}^{4/3}$ for the optical bands ($\nu_{\rm opt}<\nu_{\rm c}$) and $\propto t_{\rm obs}^{-(p-2)/2}$ for the X-ray band ($\nu_{\rm X}>\nu_{\rm m}$) for the fast cooling spectra \citep{2025MNRAS.536.1822K}. The slow cooling regime starts at $\sim 10^5$ s, which causes the lightcurve breaks as shown in Figure~\ref{fig:cornor_plot}). 

To make the first peak of the optical band $\sim 10^2$ s, the model requires bright reverse shock emission only in the optical bands. This can be achieved in fast cooling emission with a strong magnetic field. 
The onset time of the reverse shock emission is determined by the reverse shock shell crossing time, which decreases with $\sigma_0$. The high $\sigma_0$ is required 
to obtain a faster onset compared to the onset of the forward shock emission. Thus, the large $\sigma_0$ can fulfil both the requirements of the high emissivity and short shell-crossing time. 
The large $\sigma_0$ leads to initial magnetic acceleration. This is the reason for the small initial bulk Lorentz factor of the ejecta (see Fig. \ref{fig:test_GB}).

The model suggests a large initial energy, $E_0 \sim 10^{55}~{\rm erg}$ and a high electron energy fraction $\epsilon_{e, {\rm FS}} \sim 0.5$, both of which lie near the upper bounds of their prior ranges. 
Since both the $\sigma_0$ and $\Delta_0$ are significantly large to maintain the magnetic acceleration phase longer, the isotropic equivalent energy becomes large. To explain the increasing optical early afterglows, a large $\epsilon_{e,\rm FS}$ is required to make $\nu_{\rm m}$ higher than optical bands at least until $\sim 10^3$ s. 

{As we have mentioned, the jet-edge effect is required for} 
the observed decaying index of optical $(-1.80\pm0.15)$ and X-ray $(-1.98\pm0.07)$ for $10^4\ \rm{s}<t_{\rm obs}<10^6\ \rm{s}$ \citep{2011ApJ...736....7C}.  
{The lateral spreading makes the decaying index of the late afterglows further steeper after the jet break occurs \citep{1999ApJ...525..737R,2012ApJ...751..155V,2018ApJ...865...94D,2020ApJ...896..166R,2024ApJS..273...17W}.
However, hydrodynamical simulations indicate that significant lateral spreading does not occur until the jet becomes mildly relativistic \citep[e.g.][]{2012MNRAS.421..570G}.
This may compensate for our model prediction for the late afterglows without lateral spreading.}
The collimation corrected energy of the jet $E_{\rm K, jet} \sim E_0\theta_{\rm j}^2/2 \sim 5 \times 10^{50}~{\rm erg}$ is within the typical range for other GRB jets \citep{2020ApJ...900..112Z}. 

The total radiative efficiency defined as $\eta_{\rm tol}=E_{\gamma,\rm iso}/(E_{\gamma,\rm iso}+E_{0})$, is less than 1\%.
This low efficiency of the energy release in the prompt phase may be due to the high-$\sigma$ jet, in which the shock dissipation of the jet kinetic energy is inefficient.
Considering the high-sigma correction, we have $\eta=E_{\gamma,\rm iso}/(E_{\gamma,\rm iso}+E_{0}/(1+\sigma_0))=12 \pm 8\%$ which is consistent with previous studies \citep{2016MNRAS.461...51B,2022ApJ...939...39W}.

\begingroup
\renewcommand{\arraystretch}{1.3}
  \begin{table}
    \caption{
    Parameter estimation priors and marginalized posteriors for the afterglow of GRB~110213A using the \texttt{Magglow} magnetic bullet afterglow model.
    We adopt uniform priors for all parameters.
    The median values of the one-dimensional (1D) posterior distributions are reported, along with symmetric 68\% credible intervals (i.e., the 16th and 84th percentiles).
    The best-fit parameters are the values when the posterior probability is largest.
    The model assumes $\theta_{\rm obs} = 0~{\rm rad}$. 
    }
    \label{table:parameter_estimation}
    \centering
    \begin{tabular}{cccc}
      \hline
      Parameters & Prior Ranges & 1D dist. & Best-fit.\\
      \hline
      $\log(E_0~[{\rm erg}])$ & [51, 55] & $54.97^{+0.03}_{-0.11}$ & $54.98$ \\
      $\log(\Gamma_0)$  & [1.0, 3.0] & $1.62^{+0.85}_{-0.33}$ & $1.60$ \\
      $\log(\sigma_0)$ & [$-1$, 2] & $1.16^{+0.22}_{-0.54}$ & $1.22$ \\
      $\log(\Delta_0/c ~[{\rm s}])$ & [2.0, 4.0] & $3.05^{+0.04}_{-0.05}$ & $3.05$ \\
      $\log(n_0~[{\rm cm^{k-3}}])$ & [$-2.0$, 2.0] & $-0.70^{+0.24}_{-0.23}$ & $-0.81$ \\
      $k$ & [0, 2] & $0.13^{+0.18}_{-0.13}$ & $0.001$ \\
      $\log(\epsilon_{e, {\rm FS}})$ & [$-3.0$, $-0.3$] & $-0.32^{+0.02}_{-0.13}$ & $-0.31$ \\
      $\log(\epsilon_{B, {\rm FS}})$ & [$-5.0$, $-0.3$] & $-4.00^{+0.36}_{-0.23}$ & $-4.00$ \\
      $p_{\rm FS}$ & [2.01, 2.99] & $2.33^{+0.05}_{-0.04}$ & $2.32$ \\
      $\log(f_{e, {\rm FS}})$ & [$-2$, 0] & $-0.17^{+0.05}_{-0.09}$ & $-0.15$ \\
      $\log(\epsilon_{e, {\rm RS}})$ & [$-3.0$, $-0.3$] & $-1.10^{+0.64}_{-0.58}$ & $-1.12$ \\
      $p_{\rm RS}$ & [2.01, 2.99] & $2.75^{+0.23}_{-0.40}$ & $2.82$ \\
      $\log(f_{e, {\rm RS}})$ & [$-3$, 0] & $-1.26^{+0.26}_{-0.19}$ & $-1.33$ \\
      $\theta_{\rm j} ~[{\rm rad}]$ & [0.001, 1.5] & $0.01^{+0.001}_{-0.001}$ & $9.7\times10^{-3}$ \\
      \hline
      {$\chi^2/{\rm d.o.f}$} & & - & {1.94} \\
      {$\chi^2$} & & - & {710} \\
      {${\rm d.o.f}$} & & - & ${380 - 14}$ \\
      \hline
    \end{tabular}
  \end{table}
\endgroup
\begingroup

\begin{figure}
\includegraphics[width=\columnwidth]{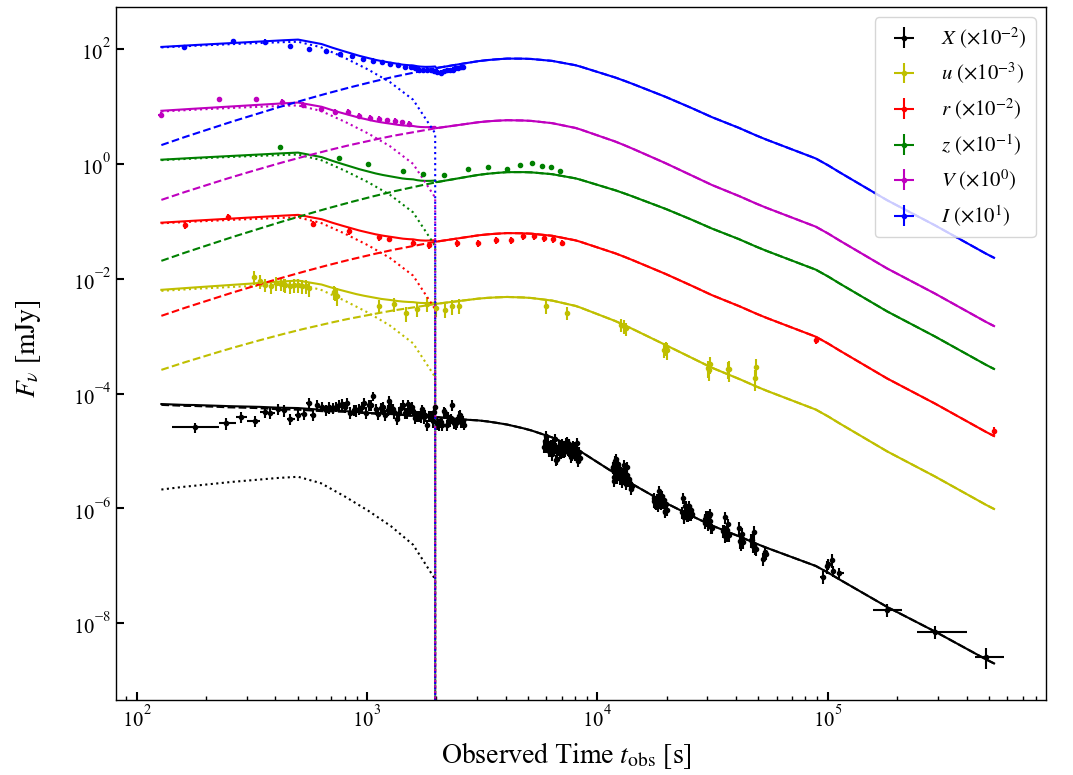}
\caption{The afterglow lightcurves of GRB~110213A in $I$-band (787 nm, blue), $V$-band (520 nm, purple), $z$-band (904 nm, green), $r$-band (624 nm, red), $u$-band (347 nm, yellow), and X-ray (10~keV, black). The data points represent extinction-corrected observations. The solid line shows the total theoretical flux, which is the sum of the emission from the forward shock (dashed line) and the reverse shock (dotted line) emission, calculated using the computational tool for the magnetic bullet afterglow \texttt{Magglow} with the best-fit parameters corresponding to the maximum posterior probability
:$E_0 = 9.5\times10^{54}~{\rm erg}$, $\Gamma_0 = 39.4$, $\sigma_0 = 16.6$, $\Delta_0/c = 1.1\times10^3$ s, $n_0 = 0.15~{\rm cm^{-3}}$, $k=0.001$, $\epsilon_{e, {\rm FS}} = 0.49$, $\epsilon_{B, {\rm FS}} = 9.9\times10^{-5}$, $p_{\rm FS}=2.32$, $f_{e, {\rm FS}}=0.71$, $\epsilon_{e, {\rm RS}} = 0.075$, $p_{\rm RS}=2.82$, $f_{r, {\rm RS}}=0.047$, and $\theta_{\rm j} = 9.7\times10^{-3}~{\rm rad}$. 
The model assumes $\theta_{\rm obs} = 0~{\rm rad}$. 
\label{fig:test_LC}}
\end{figure}

\section{Discussion} \label{sec:discuss}

Based on the best parameters obtained in the previous section, we discuss several features of the model predictions. Since the suggested CSM density structure is almost flat, we use $k=0$ in the following discussion for simplicity.

\subsection{Dynamics of the forward and reverse shocks}

\begin{figure}
\includegraphics[width=\columnwidth]{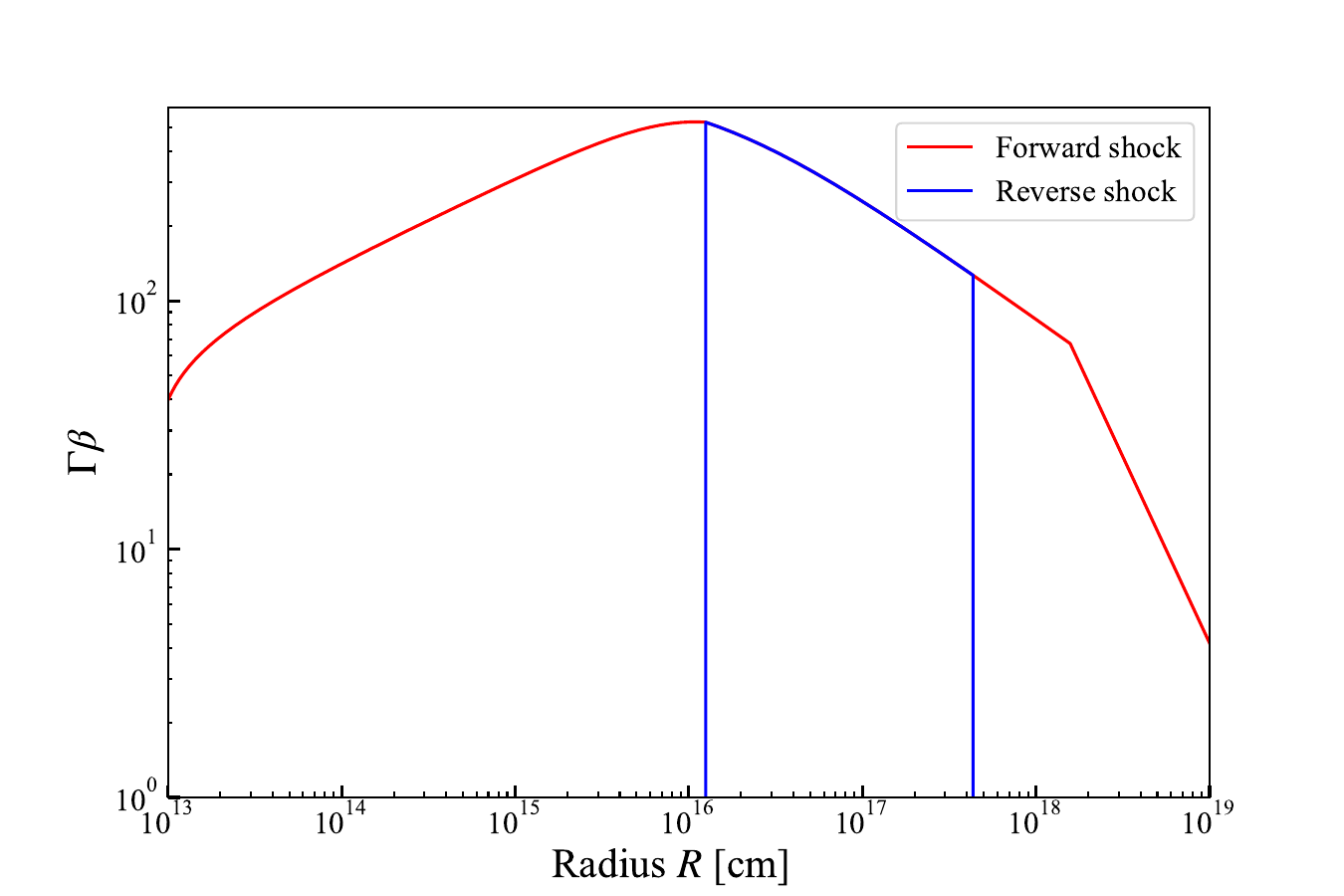}
\caption{The evolution of the Lorentz factor of the forward and reverse shocks as a function of the radius. The initial magnetic acceleration lasts until $\sim10^{16}$ cm, then the transition phase starts immediately. The deceleration phase begins after $\sim10^{18}$ cm. The reverse shock exists in $10^{16}\sim10^{18}$ cm. 
\label{fig:test_GB}}
\end{figure}

The evolution of the Lorentz factor of the forward and reverse shocks is shown in Figure \ref{fig:test_GB}. 
The jet is initially causally connected $(\theta_{\rm j}<\Gamma^{-1})$ before $\sim10^{14}$ cm, then causally disconnected by magnetic acceleration. The lateral causal connection may reduce the inhomogeneity of the internal structure of the ejecta through energy redistributions \citep{2000ApJ...535..152K}.
This may support our top-hat jet assumption.

Due to the strong magnetic pressure in the early stage of the evolution, the magnetic acceleration increases the bulk Lorentz factor $\sim500$ at $10^{16}$ cm. 
After that, the reverse shock started to propagate and cross the ejecta at $4\times10^{17}$ cm, producing the first peak in the optical light curves at $t_{\rm obs}\sim500$ s. 
The energy injection during the transition phase continues until $10^{18}$ cm due to the large $\Delta_0\gg cT_{90}$, making the second peak of the optical light curves and X-ray shallow decay. 
Finally, the forward shock decelerates following the Blandford-McKee solution. 

The estimated radial width $\Delta_0/c\approx10^3$ s is an order of magnitude larger than the burst duration of the prompt emission $cT_{90}$. 
This means that the region responsible for the prompt emission should be compact compared to the total width  $\Delta_0$. 
However, the $T_{90}$ measured in gamma-ray can be underestimated compared to which measured in X-ray \citep{2025arXiv250600435Y}.

\subsection{Spectra of the forward and reverse shocks}

\begin{figure}
\includegraphics[width=\columnwidth]{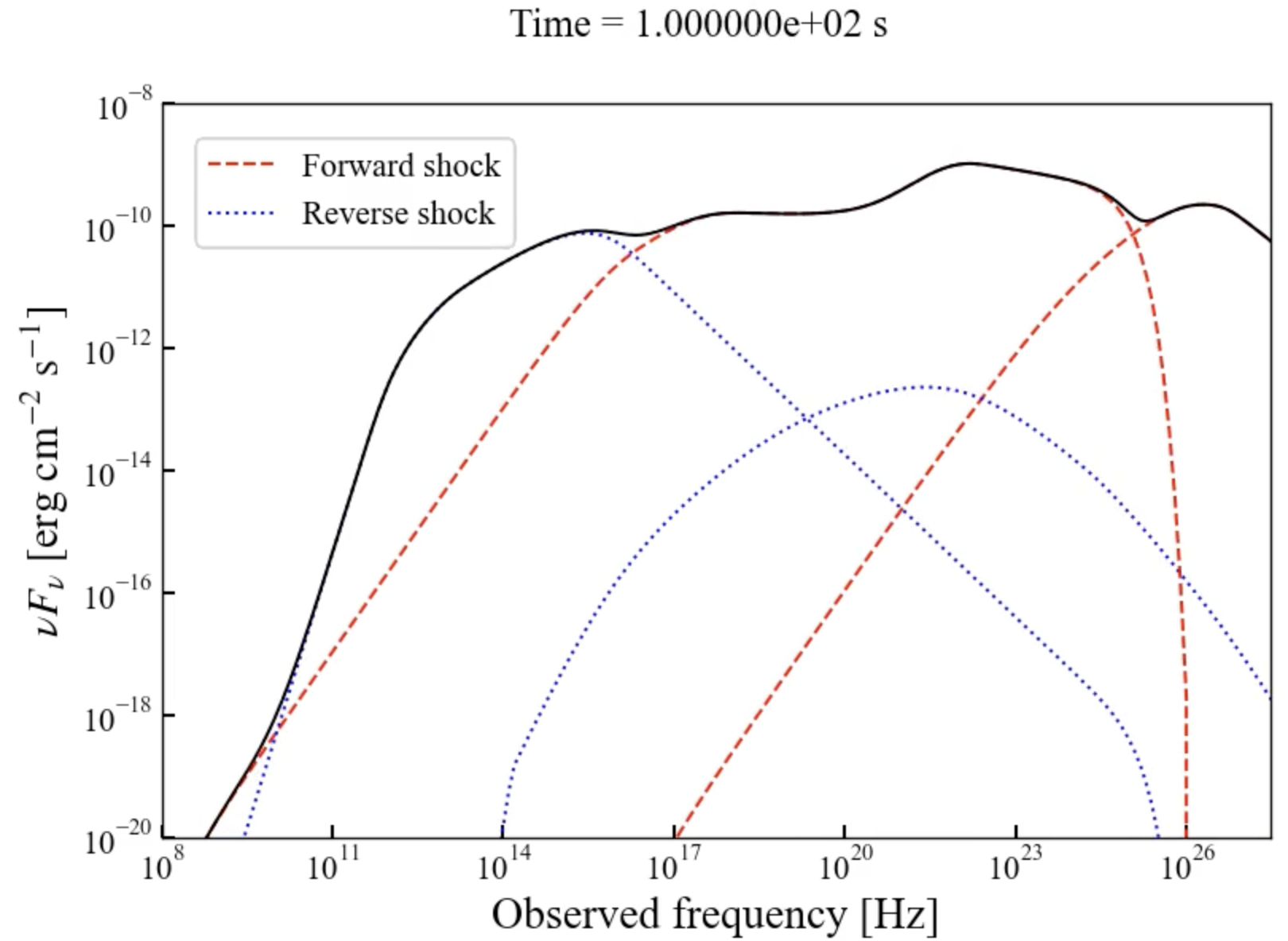}
\caption{The synthetic spectrum of synchrotron and SSC components for the forward and reverse shocks at $t_{\rm obs}=10^2$ s. 
The bump at the VHE components in the synchrotron spectrum is due to the spectral hardening of the accelerated particle distribution due to the Klein-Nishina effect. 
\label{fig:test_spec}}
\end{figure}

The synthetic spectrum at $t_{\rm obs}=10^2$ s is shown in Figure \ref{fig:test_spec}. 
Both the forward and reverse shock regions are in the fast cooling regime.
The radio emission of the reverse shock component is strongly affected by SSA. 
Because the magnetic field is strong in the reverse shock region, the SSA frequency is higher and the cooling frequency is lower than that of the forward shock component. In addition, the peak frequency is smaller because of the small $\gamma_{\rm m}$ in the reverse shock region compared to that of the forward shock. Therefore, in the optical band, the reverse shock emission is dominant.

The bump in the GeV components in the forward shock component is due to spectral hardening of the particle distribution due to the Klein-Nishina (KN) effect \citep{2009ApJ...703..675N,2010ApJ...712.1232W}. 
The spectral hardening occurs at $\hat{\nu}_{\rm m}\sim\nu_{\rm m}^{-2}\Gamma^3B$ and increases synchrotron flux by factor $\sqrt{\epsilon_e/\epsilon_B}$ compared to without the KN effect \citep{2009ApJ...703..675N}.
This specific bump due to our choice of the parameters (large $\Gamma$ and $\epsilon_e$, and small $\epsilon_B$) appears only for $t_{\rm obs}<10^4$ s.

The GeV flux is around $10^{-10}\ \rm{erg}\ \rm{cm}^{-2}\ \rm s$, which is above the Fermi-LAT sensitivity at 10 GeV $\sim10^{-11}\ \rm{erg}\ \rm{cm}^{-2}\ \rm s$ \citep{2009ApJ...697.1071A}.
The TeV flux is dominated by SSC components from the forward shock.
The TeV flux is around $10^{-10}\ \rm{erg}\ \rm{cm}^{-2}\ \rm s$.
However, considering the absorption by the extragalactic background light with its redshift (1.46), the detection with gamma-ray telescopes such as the Cherenkov Telescope Array Observatory (CTAO) is difficult.

\subsection{Early Optical and X-ray emission} \label{subsec:early_phase} 

{
The early afterglows are important tracers for determining the radial structure and magnetization of the ejecta.
Although a modelling including more realistic and diverse inner structure of the ejecta is required to fully constrain the early afterglows, it requires numerical simulations which is often computationally huge to combine with statistical fitting. 
Thus, in this study, we fix the initial density profile of the ejecta as wind-like $n_{\rm ej}(R)\propto R^{-2}$ for simplicity and have obtained well-consistent light curves. 
}

{
Our predicted light curves until the reverse shock components fading away slightly deviate from the early optical and X-ray data, which slightly increase with time. 
According to Eq. (\ref{eq:Gamma_trans}), the dynamics of the forward and reverse shocks are coupled and determined by the density ratio of the CSM and the ejecta. 
In our simple ejecta profile ($n_{\rm ej} \propto R^{-2}$), 
the density in front of the reverse shock 
is scaled as $\propto t^{-2}$ following the radial expansion. 
If the ejecta profile is steeper than we thought, the temporal index of $n_{\rm ej}$ becomes shallower than $-2$. Then, the time evolution of the forward and reverse shocks is slower than $t^{-1/2}$. 
With such a modulated density profile, the temporal index of the observed light curve can be consistent with the gradually rising X-ray light curve until $\sim 500$ s, making the reduced $\chi^2$ smaller. 
}


\subsection{Radio and gamma-ray emission}

\begin{figure}
\includegraphics[width=\columnwidth]{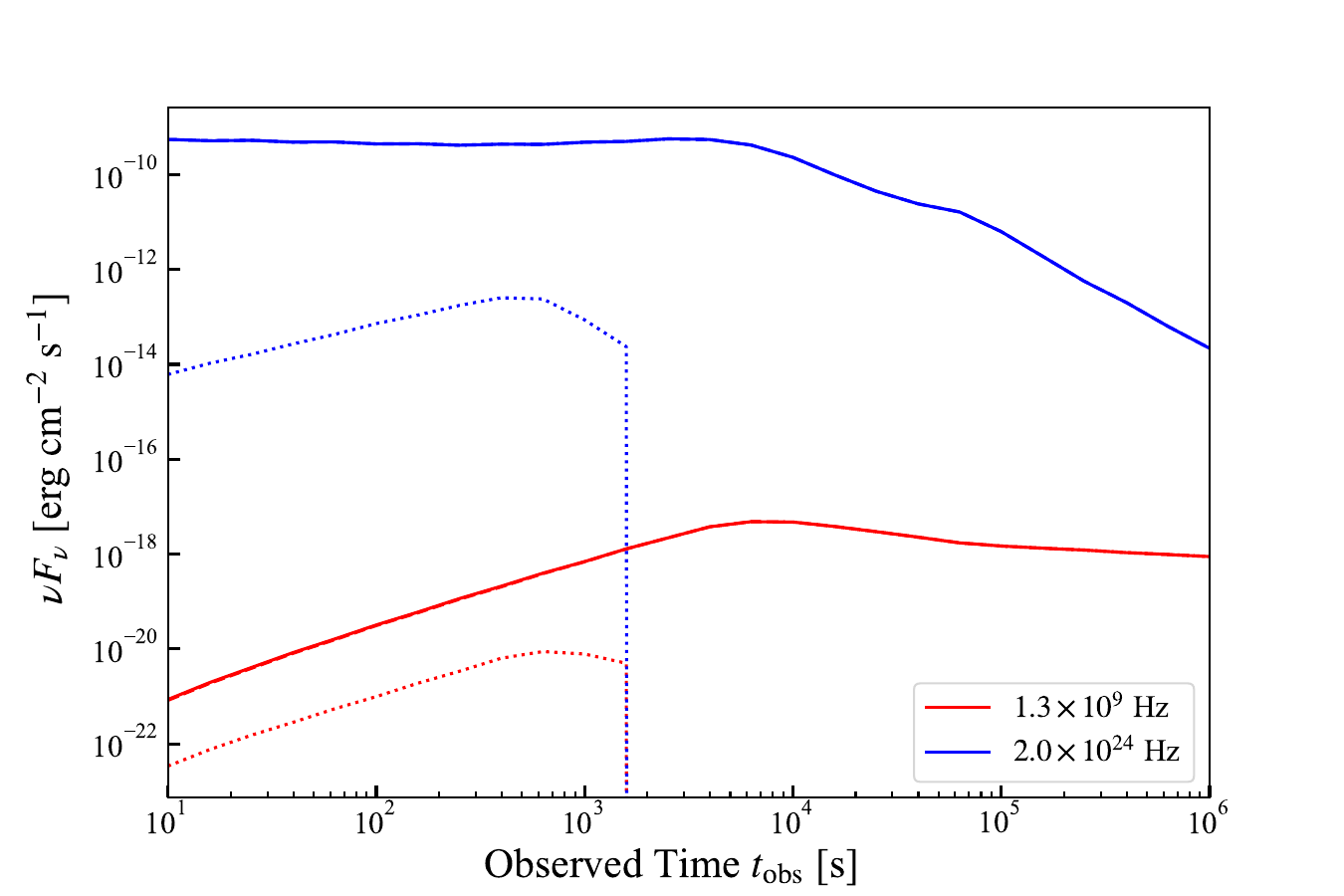}
\caption{The predicted lightcurves of the radio and GeV afterglow. The solid line shows the total theoretical flux, which is the sum of the emission from the forward shock (dashed line) and the reverse shock (dotted line) emission, calculated using the magnetic bullet afterglow model \texttt{Magglow} with the best-fit parameters corresponding to the maximum posterior probability.
\label{fig:LC_other}}
\end{figure}

The follow-up observation of radio afterglows after the jet break will constrain the late-phase dynamics of the Magnetic Bullet model.
The lightcurves of radio and gamma-ray afterglow based on our derived best-fitting parameters are shown in Figure \ref{fig:LC_other}. 
The forward shock components are brighter than the reverse shock components at any time. 
Although the flux level of radio afterglow is above the sensitivity of the Very Large Array (VLA) $\sim10^{-19}\ \rm{erg\ cm^{-2}\ s^{-1}}$ at 1 GHz, 
no observations have been reported so far \citep{2012ApJ...746..156C}. 
The late-phase radio emission may be highly sensitive to lateral spreading \citep{2020ApJ...896..166R}, which we do not yet include.  

Our predicted lightcurves suggest that the GeV afterglow also shows shallow decaying features. 
The high VHE flux is due to the spectral hardening of accelerated electrons by the Klein-Nishina effect, which leads to the VHE bump in the synchrotron spectra as shown in Figure \ref{fig:test_spec}. 
Although the flux level is above the Fermi-LAT sensitivity, no observation was performed because the angle from the Fermi-LAT boresight is 101 degrees \citep{2011GCN.11727....1F}. 
Thus, simultaneous detections of both X-ray and GeV gamma-ray shallow decay will be expected by future rapid follow-up observations such as CTAO. 

\subsection{Comparison with other GRBs}

Our best-fitting parameters suggest that GRB 110213A has a large isotropic equivalent energy $E_{\rm iso}\approx10^{55}$ erg with an exceptionally narrow opening angle $\theta_{\rm j}\approx0.01$ rad. 
A similar trend is reported in a couple of GRBs. 
\citet{2024JHEAp..41....1O} point out that GRB 0808710 has $E_{\rm iso}\approx10^{55}$ erg with $\theta_{\rm j}\approx0.01$ rad. They also reported the estimated radiative efficiency is much less than $1\%$, which is similar to the case of GRB 110213A. 
However, their used model shows a discrepancy in the shallow rising phase $\propto t_{\rm obs}^{1.1}$ of the early optical lightcurves \citep{2009A&A...508..593K}. Because this temporal index is similar to our result (see dashed lines in Figure \ref{fig:test_LC}), the Magnetic Bullet model may be suitable to describe the entire light curves of GRB 080710. 

\citet{2025MNRAS.540.2098K} show that GRB 221009A has $E_{\rm iso}\approx4\times10^{55}$ erg with $\theta_{\rm j}\approx0.03$ rad. 
They claim that the ejecta should be strongly magnetized to accelerate up to $\Gamma_{\rm max}\approx500$ at $R\sim10^{16}$ cm, which is also similar for the case of GRB 110213A (see Figure \ref{fig:test_GB}). However, the apparent radiative efficiency is very small for GRB 110213A compared to $\sim10\%$ for the case of GRB 221009A. 
This suggests that efficient magnetic energy dissipation occurred in the prompt emission for GRB 221009A, but not for GRB 110213A. 
The magnetic dissipation efficiency is almost negligible for the case of internal shock within the strongly magnetized ejecta \citep{1984ApJ...283..694K}.
However, \citet{2023MNRAS.526..512K} have demonstrated that the intermittent multiple shock interactions of strongly magnetized ejecta can dissipate their magnetic energy $\sim 10\%$. 
The difference in radiative efficiency of several GRBs may decipher the physics of the jet-launching mechanism.

\section{Conclusions} \label{sec:summary}

In this paper, we studied the origins of the double-peaked optical light curves with X-ray shallow decay for GRB 110213A.
The early afterglows are important tracers for determining the structure and magnetization of the ejecta. However, no comprehensive model has been created so far. Thanks to advancing the recent numerical simulation for strongly magnetized relativistic jets, we construct the unified semi-analytic model for describing the forward and reverse shock dynamics driven by an interaction between magnetized ejecta and stratified CSM. We also implemented the synthetic spectra and the light curve calculation tool as the open-source Julia package \texttt{Magglow}. Using the MultiNest nested sampling algorithm, we applied our model to constrain the several parameters for GRB 110213A.

Our results suggest that the double-peaked optical light curves are obtained considering strongly magnetized thick shell ejecta. 
The first peak corresponds to the reverse shock emission in the magnetized ejecta, while the second peak is dominated by the forward shock emission. 
The shallow decay of the X-ray light curve is created through the suppression of the forward shock deceleration in the transition phase.
Inferred thickness of the ejecta $\Delta_0$ is an order of magnitude larger than $cT_{90}$, which gives us a theoretical challenge of connecting the mechanisms of prompt and afterglow emissions. In fact, the large energy $E_{\rm iso}$ and magnetization $\sigma$ of the ejecta are required to reproduce the light curves. The estimated low radiative efficiency may be a consequence of the high value of $\sigma$. 
Although no follow-up observation was made for radio and GeV gamma-ray for GRB 110213A, our predicted light curves are well above the sensitivity of current and future telescopes. 
We expect that analysis of multiwavelength early afterglows by our simulator \texttt{Magglow} will decipher the inner structure of relativistic jets.

\section*{Acknowledgements}

The authors thank anonymous referee for thoughtful discussions on this work. 
The authors thankfully acknowledge the computer resources provided by the Institute for Cosmic Ray Research (ICRR), the University of Tokyo.
K.O. thanks Dr.~K.~Murase for his guidance on the use of MultiNest.
This work is supported by the joint research program of ICRR, and JSPS KAKENHI Grant Numbers JP23KJ0692 (Y.K.), 22K03684, 23H04899, 24H00025, 25K07352 (K.A.), 23K22522, 23K25907 (R.Y.), 
and by JST SPRING, Grant Number JPMJSP2103 (K.O.). 

\section*{Data Availability}

The code presented and used in this work is open source and can be found at \url{https://github.com/yo3-sun/Magglow}. 
The data underlying this article will be shared on reasonable request to the corresponding author.

\begin{figure*}
\includegraphics[width=\linewidth]{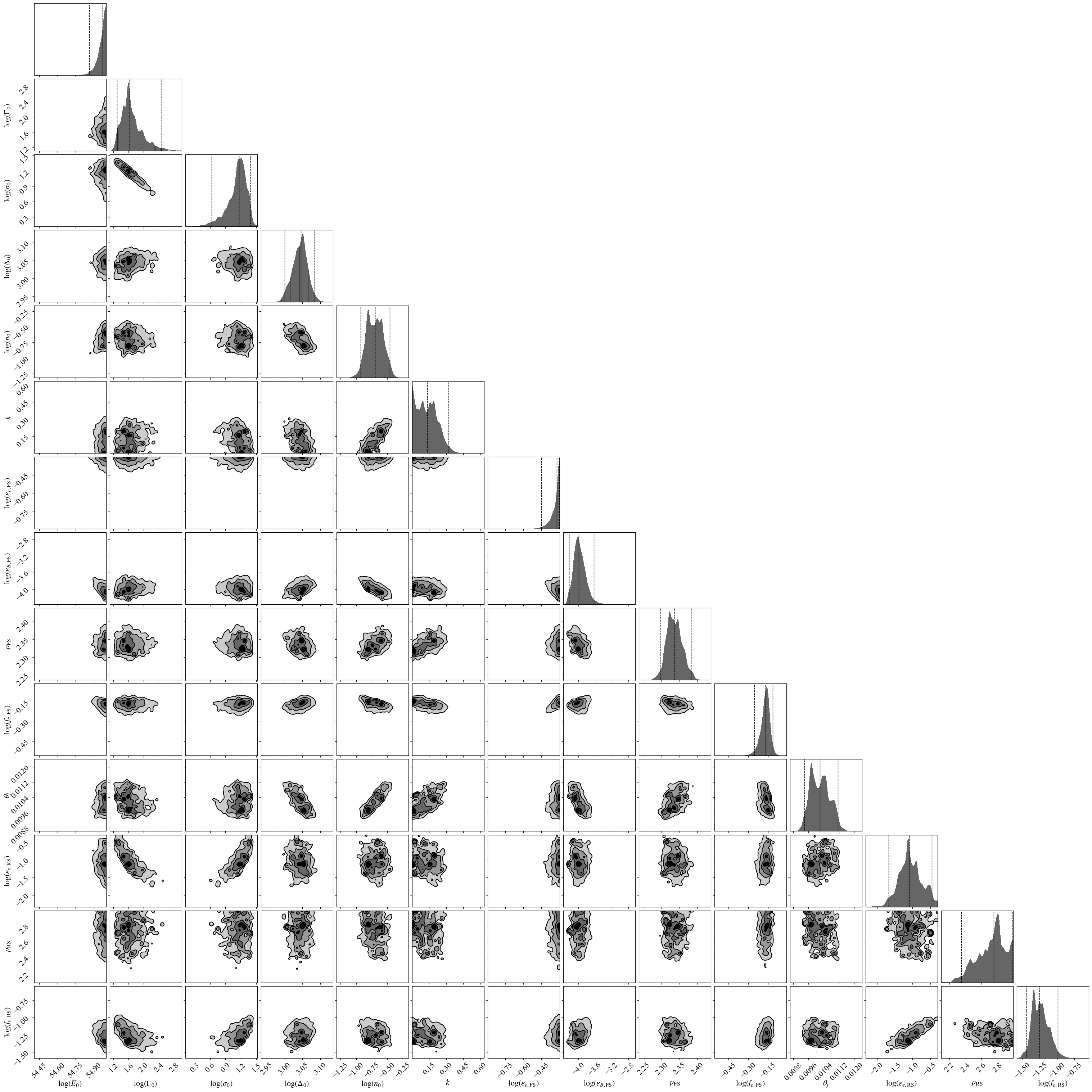}
\caption{Corner plot showing the posterior probability distributions of the model parameters for the GRB~110213A afterglow, obtained using the magnetic bullet afterglow model.
Two-dimensional projections of the parameter samples are shown to highlight correlations and covariances between parameters.
The uncertainties correspond to the 16th and 84th percentiles of the marginalized posterior distributions, representing the 1$\sigma$ credible intervals, and are indicated by black dashed lines.
Flat priors were adopted for all parameters in the multimodal nested sampling analysis.
\label{fig:cornor_plot}}
\end{figure*}



\bibliographystyle{mnras}
\bibliography{main} 




\appendix

\bsp	
\label{lastpage}
\end{document}